\documentclass[aps,twocolumn,prl]{revtex4-2}

\usepackage{amsmath,graphicx,color}
\usepackage{bm}
\usepackage{physics}

\newcommand{\pd}{{\phantom{\dagger}}}

\newcommand{\FST}{FeSe$_{0.45}$Te$_{0.55}$ }
\newcommand{\FSTT}{FeSe$_{0.45}$Te$_{0.55}$}

\begin{document}

\title{Shot-noise and differential conductance as signatures of putative topological superconductivity in FeSe$_{0.45}$Te$_{0.55}$}

\author{Ka Ho Wong$^{1}$, Eric Mascot$^{1}$, Vidya Madhavan$^{2}$, Dale J. Van Harlingen$^{2}$, and Dirk K. Morr$^{1}$}
\affiliation{$^{1}$ University of Illinois at Chicago, Chicago, IL 60607, USA \\
$^{2}$ University of Illinois at Urbana Champaign, Champaign, IL 61801, USA}

\date{\today}

\begin{abstract}
We present a theory for the differential shot noise, $dS/dV$, as measured via shot-noise scanning tunneling spectroscopy, and the differential conductance, $dI/dV$, for tunneling into Majorana zero modes (MZMs) in the putative topological superconductor FeSe$_{0.45}$Te$_{0.55}$. We demonstrate that for tunneling into chiral Majorana edge modes near domain walls, as well as MZMs localized in vortex cores and at the end of defect lines, $dS/dV$ vanishes whenever $dI/dV$ reaches a quantized value proportional to the quantum of conductance. These results are independent of the particular orbital tunneling path, thus establishing a vanishing $dS/dV$ concomitant with a quantized $dI/dV$, as universal signatures for Majorana modes in two-dimensional topological superconductors, irrespective of the material's specific complex electronic bandstructure.
\end{abstract}

\maketitle

{\it Introduction} The unambiguous experimental identification of Majorana zero modes (MZMs) in topological superconductors, the putative building blocks for topological quantum computing, has remained a major challenge. While it has been predicted that the differential conductance, $dI/dV$, for tunneling into MZMs is quantized both for one-dimensional \cite{Law2009,Flensberg2010} and two-dimensional topological superconductors \cite{Rachel2017}, and that the shot-noise associated with the tunneling into Majorana modes exhibits various characteristic signatures \cite{Bolech2007,Law2009,Zocher2013,Diez2014,Liu2015,Jonckheere2019,Beenakker2020,Perrin2020}, the experimental verification of these predictions has remained difficult due to the close proximity of trivial in-gap states, as well as small superconducting gaps. Recent experiments, however, have provided enthralling evidence for the existence of topological surface superconductivity in the iron-based superconductor FeSe$_{0.45}$Te$_{0.55}$  possessing a full $s_\pm$-wave gap of a few meV. In particular, these experiments have reported the existence of a surface Dirac cone \cite{Zhang2018}, of MZMs in the vortex core \cite{Wang2018,Machida2019,Zhu2020} and at the end of line defects \cite{Chen2020} in monolayer \FST, and of a Majorana edge mode at a domain wall \cite{Wang2020}. While the physical origin of this putative topological phase has remained a question of debate, either being ascribed to the existence of a topological insulator whose surface Dirac cone is gapped out by proximitized superconductivity \cite{Wang2015,Zhang2018}, or to the interplay \cite{Mascot2021} of an $s_\pm$-wave gap, a Rashba spin-orbit interaction and recently observed surface ferromagnetism \cite{Zaki2019,Du2021}, the question naturally arises of what type of universal physical observables associated with the existence of localized MZMs or chiral Majorana edge modes can be expected in the multi-orbital FeSe$_{0.45}$Te$_{0.55}$ compound.

The development of scanning tunneling shot-noise spectroscopy \cite{Birk1995,Kemiktarak2007,Herz2013,Bastiaans2018,Massee2019} has opened a new approach to investigating this question as it allows the direct measurement of the local shot-noise associated with electron tunneling into a Majorana mode, thus complementing the measurement of the differential conductance. Using a Keldysh Greens function approach, we compute the differential shot noise, $dS/dV$, and differential conductance, $dI/dV$, using a recently proposed 5-orbital model for FeSe$_{0.45}$Te$_{0.55}$ in which the topological superconducting phase arises from the interplay of surface magnetism, a Rashba spin-orbit interaction, and a hard superconducting gap with $s_\pm$-symmetry \cite{Mascot2021}. Specifically, we study $dS/dV$ and $dI/dV$ for three different occurrences of Majorana modes: a chiral Majorana edge mode at a domain wall, and MZMs in vortex cores and at the end of line defects. We demonstrate that for each of these cases, the electron tunneling into a Majorana mode is accompanied by a vanishing differential shot noise, $dS/dV$, and a quantized differential conductance, $dI/dV$, being equal to the quantum of conductance.  Since these results hold for tunneling into each of the 5 relevant orbitals, they demonstrate that a quantized $dI/dV$, and a vanishing $dS/dV$ are universal signatures for Majorana modes in two-dimensional topological superconductors, that are independent of the material's specific complex electronic bandstructure. This, in turn, allows one to employ the vanishing of the differential shot noise as a local marker to detect topological phase transitions. Finally, we show that the measurement of a non-vanishing supercurrent along a domain wall using a Scanning SQUID (superconducting quantum interference device) Microscope (SSM) \cite{Spanton2014} is a third characteristic feature for the existence of Majorana modes, thus complementing those features found in $dS/dV$ and $dI/dV$. The combination of our results points toward new possibilities for the experimental identification of Majorana modes in complex electronic materials.

\noindent {\it Theoretical Model} To investigate the shot noise and differential conductance associated with the tunneling into Majorana modes on the surface of \FSTT, we consider a two-dimensional 5-orbital model \cite{Gra09,Sarkar2017} that was obtained from a fit to ARPES and STS experiments \cite{Sarkar2017} and was recently proposed to explain the emergence of topological superconductivity in \FST as arising from the interplay between (i) a full superconducting $s_\pm$-wave gap, (ii) surface magnetism, evidence for which was recently reported by ARPES \cite{Zaki2019} and quantum sensing \cite{Du2021} experiments, and (iii) a Rashba spin orbit (RSO) interaction that arises from the breaking of the inversion symmetry on the surface. The resulting Hamiltonian in real space is given by
\begin{align}
 H_{0} =& -\sum_{a,b=1}^{5} \sum_{{\bf r},{\bf r'},\sigma} t^{ab}_{\bf r,r'} c_{{\bf r},a,\sigma}^{\dagger} c_{{\bf r'},b,\sigma} - \sum_{a=1}^{5} \sum_{{\bf r},\sigma} \mu_{aa}  c_{{\bf r},a,\sigma}^{\dagger} c_{{\bf r},a,\sigma} \nonumber \\
 & +i \alpha \sum_{a=1}^{5} \sum_{{\bf r}, {\bm \delta}, \sigma, \sigma'} c_{{\bf r},a,\sigma}^\dag \left( {\bm \delta}
\times {\bm \sigma}\right)^z_{\sigma \sigma'} c^\pd_{{\bf r}+{\bm \delta},a, \sigma'} \nonumber \\
 & - J  \sum_{a=1}^{5} \sum_{{\bf r},\sigma, \sigma'} {\bf S}_{\bf r} \cdot c_{{\bf r},a, \sigma}^\dag {\boldsymbol \sigma}_{\sigma \sigma'} c^\pd_{{\bf r},a, \sigma'}  \nonumber \\
& + \sum_{a=1}^{5}  \sum_{\langle \langle {\bf r},{\bf r'} \rangle \rangle} \Delta^{aa}_{{\bf r}{\bf r}'} c_{{\bf r},a,\uparrow}^{\dagger} c_{{\bf r'},a,\downarrow}^{\dagger} + {\rm H.c.}
 \label{eq:Hamiltonian}
\end{align}
Here $a,b=1,...,5$ are the orbital indices corresponding to the $d_{xz}$-, $d_{yz}$-, $d_{x^2-y^2}$-, $d_{xy}$-, and $d_{3z^2-r^2}$-orbitals, respectively,  $-t^{ab}_{\bf rr'} $ represents the electronic hopping amplitude between orbital $a$ at site ${\bf r}$ and orbital $b$ at site ${\bf r'}$ on a two-dimensional square lattice, $\mu_{aa}$ is the on-site energy in orbital $a$, $c_{{\bf r},a,\sigma}^{\dagger} (c_{{\bf r},a,\sigma})$ creates (annihilates) an electron with spin $\sigma$ at site ${\bf r}$ in orbital $a$, and ${\bm \sigma}$ is the vector of spin Pauli matrices. The superconducting order parameter $\Delta^{aa}_{{\bf r}{\bf r}'}$ represents intra-orbital pairing between next-nearest neighbor Fe sites ${\bf r}$ and ${\bf r}'$ (in the 1 Fe unit cell), yielding a superconducting $s_\pm$-wave symmetry \cite{Sarkar2017}. Moreover, $\alpha$ represents the Rashba spin-orbit interaction arising from the breaking of the inversion symmetry at the surface\,\cite{NadjPerge2014} with $\bm \delta$ being the vector connecting nearest neighbor sites. ${\bf S}_{\bf r}$ denotes the ferromagnetically ordered moment \cite{Zaki2019, Du2021} which is locally exchanged coupled to the conduction electrons via an interaction $J$. The observation of ARPES experiments \cite{Zaki2019} are consistent with a considerable fraction of the ordered magnetic moment aligned perpendicular to the surface, such that we assume an out-of-plane ferromagnetic alignment of ${\bf S}_{\bf r}$ for concreteness. In the normal state, the Hamiltonian of Eq.(\ref{eq:Hamiltonian}) yields three Fermi surfaces in the 1Fe Brillouin zone whose orbital character implies that the superconducting order parameter is non-zero only  in the $d_{xz}$-, {$d_{yz}$-}, and $d_{xy}$-orbitals \cite{Sarkar2017}.
Finally, as shown in Ref.~\cite{Mascot2021}, this model exhibits a series of topological phases that are characterized by the Chern number, $C$.

To compute the differential conductance and differential shot noise in a topological superconducting phase, we employ the Keldysh Greens function formalism, and obtain for the current flowing between the tip and the system
\begin{align}
I(V)=\frac{2 e t_0}{h}\sum_{\sigma=\uparrow,\downarrow}\int_{-eV}^{eV}\dd{\varepsilon}\mathrm{Re}\left[G^{<}_{ts}\left(\vb{r},\sigma,\sigma,\varepsilon\right)\right] \ ,
\label{eq:IV}
\end{align}
and for the zero-frequency shot-noise
\begin{align}
S(\omega&=0,V)=\frac{2e^2 t_0^2}{h}	\sum_{\sigma,\sigma'=\uparrow,\downarrow}\int_{-eV}^{eV}\dd{\varepsilon} \nonumber \\
& \hspace{-1.cm} \times \big[G_{tt}^{>}\qty(\sigma,\sigma',\varepsilon)G^{<}_{ss}\qty(\sigma',\sigma,\varepsilon)
+G_{ss}^{>}\qty(\sigma,\sigma',\varepsilon)G^{<}_{tt}\qty(\sigma',\sigma,\varepsilon) \nonumber \\
& \hspace{-1.cm} +[F^{>}_{ss}\qty(\sigma,\sigma',\varepsilon)]^*F^{<}_{tt}\qty(\sigma',\sigma,\varepsilon)+[F^{>}_{tt}\qty(\sigma,\sigma',\varepsilon)]^*F^{<}_{ss}\qty(\sigma',\sigma,\varepsilon) \nonumber\\
& \hspace{-1.cm}-G^{>}_{ts}\qty(\sigma,\sigma',\varepsilon)G^{<}_{ts}\qty(\sigma',\sigma,\varepsilon)-G^{>}_{st}\qty(\sigma,\sigma',\varepsilon)G^{<}_{st}\qty(\sigma',\sigma,\varepsilon) \nonumber\\
& \hspace{-1.cm} -[F^{>}_{st}\qty(\sigma,\sigma',\varepsilon)]^*F^{<}_{st}\qty(\sigma',\sigma,\varepsilon)-[F^>_{ts}\qty(\sigma,\sigma',\varepsilon)]^*F^{<}_{ts}\qty(\sigma',\sigma,\varepsilon)\big]	
\label{eq:SV}
\end{align}
Here, $G,F$ are the normal and anomalous Greens functions, with $t,s$ denoting the sites of the tip and the system between which electrons tunnel, and all Green's functions involving the system are evaluated at site ${\bf r}$ and for orbital $a$ [for details, see Supplemental Material (SM) Sec.~1]. $dI/dV$ and $dS/dV$ are then obtained by differentiating Eqs.(\ref{eq:IV}) and (\ref{eq:SV}), respectively.

\begin{figure}[t]
\centering
\includegraphics[width=8.5cm]{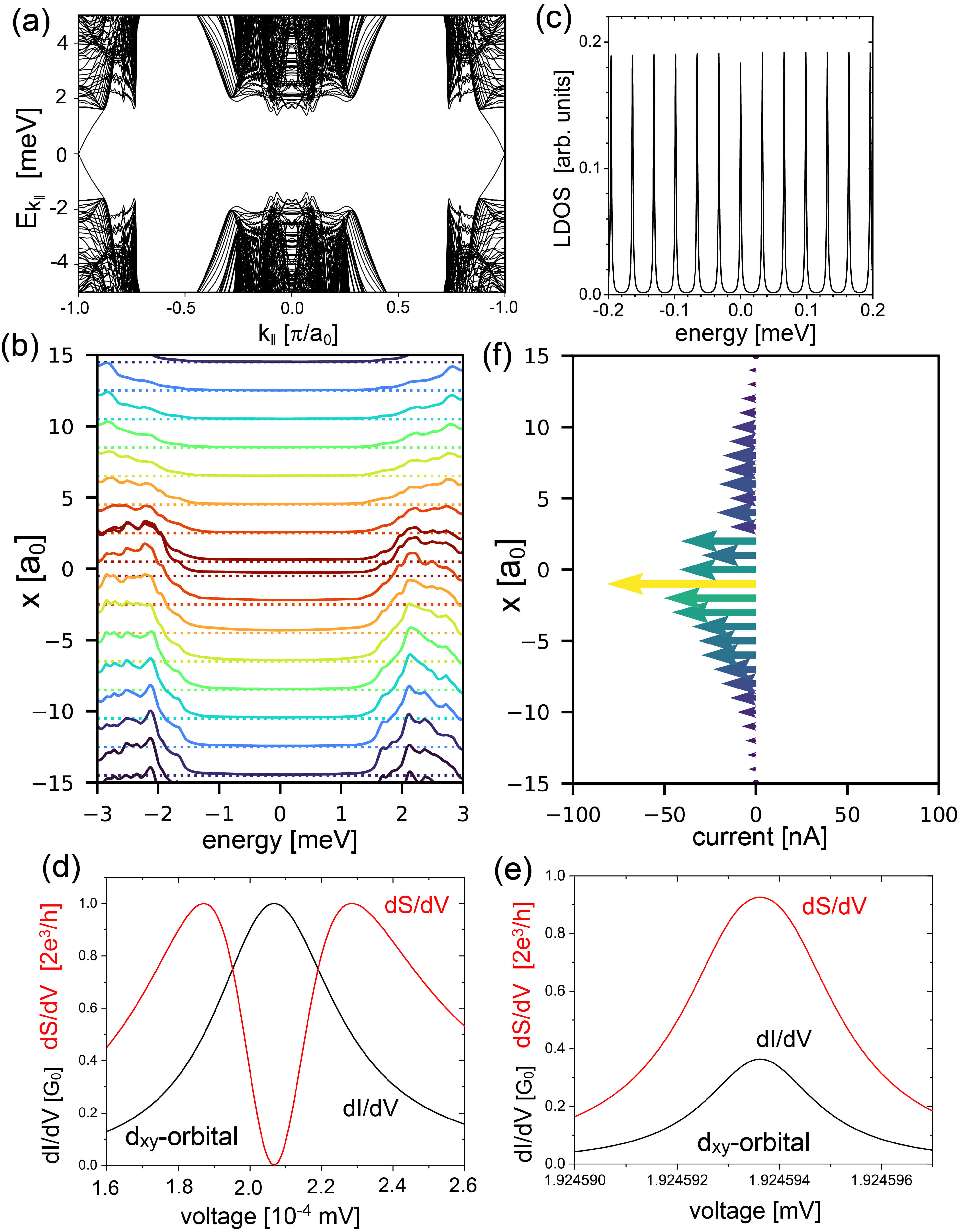}
\caption{Domain wall with $\alpha=7$ meV separating a topological $C=1$ regions with $J=7.5$ meV from a trivial $C=0$ region with $J=0$. (a) Electronic dispersion as a function of momentum along the domain wall. The chiral Majorana edge mode traverses the superconducting gap. (b) LDOS along a linecut perpendicular to the domain wall. (c) Low energy LDOS in which individual Majorana states are energetically resolved. $dS/dV$ (red line) and $dI/dV$ (black line) for (d) the lowest energy Majorana state, and (e) for a state near the gap edge with $E=1.9245$. (f) Spatial profile of the supercurrent near the domain wall. }
\label{fig:Fig1}
\end{figure}

{\it Results} There are three distinct cases in which Majorana modes emerge from the above described theoretical model for \FST: chiral Majorana edge modes along domain walls, MZMs located in vortex cores, as well as MZMs located at the end of line defects \cite{Mascot2021}. We begin by studying the form of $dS/dV$ and $dI/dV$ for chiral Majorana edge modes that emerge at a domain wall separating a topological $C=1$ region from a trivial $C=0$ region. The observation of strong disorder in \FST \cite{Cho2019,Wang2021}, combined with the fact that MZMs are not observed in all vortex cores, suggest that such domain walls could be realized on the surface of \FSTT. In Fig.~\ref{fig:Fig1}(a), we present the electronic dispersion as a function of momentum along the domain wall, showing two Majorana modes traversing the superconducting gap. We note that as we consider a system with periodic boundary conditions, the system considered here contains two domain walls, with a Chern number change of $\Delta C=1$ at each of the domain walls, resulting in two Majorana modes, one located at each of the domain walls.
%%%%%%%%%%%%%%%%%%%%%%%%%%%%%%%%%%%%%%%%%%%%%
%                  F I G .     1
%%%%%%%%%%%%%%%%%%%%%%%%%%%%%%%%%%%%%%%%%%%%%
In Fig.~\ref{fig:Fig1}(b), we show a linecut of the local density of states (LDOS) across the domain wall: it shows an almost energy independent spectral weight inside the superconducting gap near the domain wall, arising from the chiral Majorana edge mode. In order to identify the universal features of the differential shot-noise and differential conductance associated with the tunneling into a single Majorana state of this chiral edge mode, however, we need to (a) resolve the Majorana states in energy, which can be done by considering a finite length of the domain wall, leading to discretized energy levels as revealed by the LDOS shown in Fig.~\ref{fig:Fig1}(c), and (b) ensure that a finite lifetime of the Majorana states arises solely from its coupling to the tip. In Fig.~\ref{fig:Fig1}(d), we present the voltage dependence of both $dS/dV$ and $dI/dV$ near the lowest energy Majorana state, for tunneling into the $d_{xy}$-orbital. Note that due to the finite size of the system we consider, the lowest energy Majorana state is located at a very small, but finite energy.  As previously reported, we find that $dI/dV$ reaches the quantum of conductance, $G_0$, at the energy of the Majorana state \cite{Rachel2017}, implying that at this energy, the transmission amplitude reaches unity. A qualitatively new feature of the Majorana modes is revealed by the differential shot noise, dS/dV, which vanishes at that bias where $dI/dV$ reaches $G_0$. The same result holds for tunneling into any of the other $d$-orbitals that possess spectral weight at the domain wall (see SM Sec.~2), implying that a quantized conductance and vanishing differential noise are universal features that are independent of the complex multi-orbital structure of a topological superconductor. We note that by defining a transmission amplitude $T(V)$ via $dI/dV = G_0 T(V)$, the differential shot noise obeys $dS/dV \sim T(V)[1-T(V)]$ (see SM Sec.~2). A quantized $dI/dV$ and vanishing $dS/dV$ are also found for higher energy Majorana states, until one approaches states near the gap edge. Indeed, for states close to the gap edge, we find significant deviations of $dI/dV$ from $G_0$, while $dS/dV$ does not vanish any longer, as shown in Fig.~\ref{fig:Fig1}(e). This demonstrates that both a quantized value for $dI/dV$ and the vanishing of $dS/dV$ are hallmarks of chiral Majorana edge modes. Concomitant with these results, we find that the domain wall carrying a chiral Majorana edge modes also gives rise to a non-vanishing supercurrent that flows parallel to the domain wall, as shown in Fig.~\ref{fig:Fig1}(f). Such a supercurrent can be detected using an SSM scanned along the domain wall \cite{Spanton2014}, such that the combination of $dI/dV$, $dS/dV$ and SSM measurements can provide strong evidence for the existence of Majorana modes along a domain wall.

\begin{figure}[t]
\centering
\includegraphics[width=8.5cm]{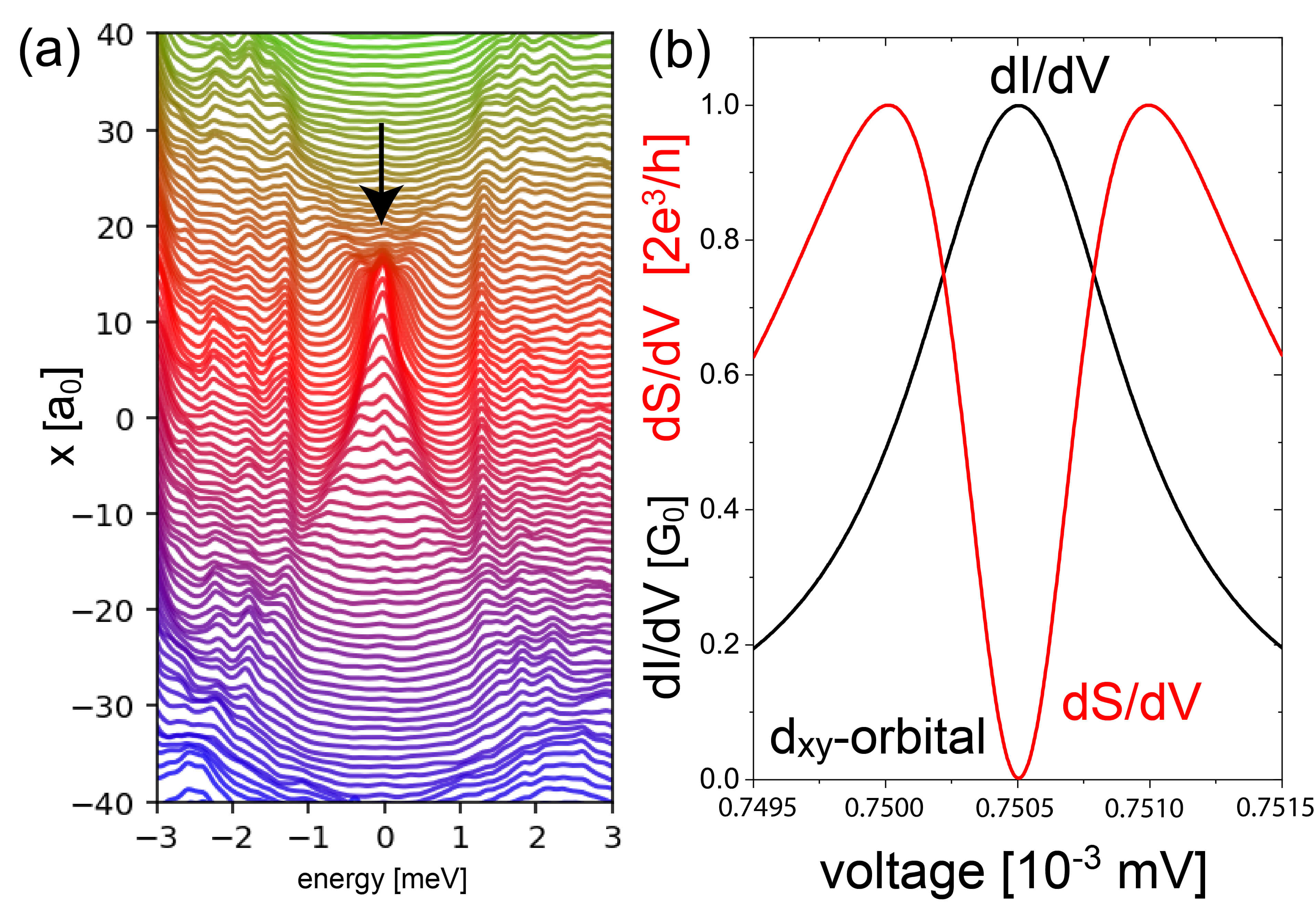}
\caption{(a) Linecut of the LDOS through the center of a vortex core in the topological $C=1$ phase with $(\alpha,J)=(7,8)$ meV for a uniform magnetic flux of $1$ Wb, revealing the existence of a MZM. (b) $dI/dV$ (black line) and $dS/dV$ (red line) for tunneling into the MZM for the $d_{xy}$-orbital.}
\label{fig:Fig2}
\end{figure}

We next study the differential shot noise associated with the tunneling into a localized MZM in a vortex core by implementing the magnetic field via the Peierls substitution and compute the spatial dependence of the superconducting order parameters in the $d_{xz}$-, $d_{yz}$-, and $d_{xy}$-orbitals self-consistently (for details, see Ref.~\cite{Mascot2021}). In Fig.~\ref{fig:Fig2}(a) we plot a linecut of the LDOS through the center of the vortex core in the $C=1$ phase, which reveals the existence of a MZM in its center (see black arrow). In Fig.~\ref{fig:Fig2}(b), we present $dS/dV$ and $dI/dV$, associated with the tunneling into this localized MZM.
%%%%%%%%%%%%%%%%%%%%%%%%%%%%%%%%%%%%%%%%%%%%%
%                  F I G .     2
%%%%%%%%%%%%%%%%%%%%%%%%%%%%%%%%%%%%%%%%%%%%%
As expected, we again find that at the bias when $dI/dV$ reaches the quantized quantum of conductance, $dS/dV$ vanishes. As shown in SM Sec.~3, this results holds for tunneling into any of the $d$-orbitals, again supporting our conclusion that these results are universal and independent of any complex electronic bandstructure.

Finally, we consider the emergence of zero energy states at the end of line defects, recently observed in monolayer FeSe$_{0.5}$Te$_{0.5}$ deposited on a SrTiO$_3$ substrate \cite{Chen2020}, whose spatial structure is similar to that expected for MZMs. It is presently unclear whether these MZMs are a characteristic feature of an underlying topological phase as argued in Ref.~\cite{Mascot2021}, similar to line defect MZMs predicted to occur in topological $p_x + ip_y$-wave superconductors \cite{Wimmer2010}, or are independent of it, simply utilizing the monolayer's complex electronic structure to form a 1D topological superconductor as proposed in Refs.\cite{Chen2020,Zhang2021,Wu2020}. Indeed, it is presently unknown whether the FeSe$_{0.5}$Te$_{0.5}$/SrTiO$_3$ system itself is topological or not. While the mechanism proposed in Refs.~\cite{Wang2015,Zhang2018} requires a 3D bulk structure, and thus would not apply to this system, the mechanism described by Eq.(\ref{eq:Hamiltonian}) could give rise to topological superconductivity if FeSe$_{0.5}$Te$_{0.5}$/SrTiO$_3$ were also to exhibit ferromagnetism (which is presently unknown). Here, we follow the argumentation of Ref.~\cite{Mascot2021}, which has demonstrated that the emergence of MZMs at the end of line defects is a characteristic feature of the underlying topological phase. To study $dS/dV$ and $dI/dV$ associated with these line defect MZMs, we represent the line defect for simplicity as a line of potential scatterers (though magnetic scatterers could also be realized \cite{Wu2020,Zhang2021,Wu2020}) described by the Hamiltonian
\begin{align}
H_{def} = U_0 \sum_{a=1}^{5} \sum_{{\bf R},\sigma} c_{{\bf R},a,\sigma}^{\dagger} c_{{\bf R},a,\sigma} \ ,
\end{align}
where $U_0$ is the potential scattering strength, and the sum runs over all sites ${\bf R}$ of the line defect.
%%%%%%%%%%%%%%%%%%%%%%%%%%%%%%%%%%%%%%%%%%%%%
%                  F I G .     3
%%%%%%%%%%%%%%%%%%%%%%%%%%%%%%%%%%%%%%%%%%%%%
\begin{figure}[t]
\centering
\includegraphics[width=8.5cm]{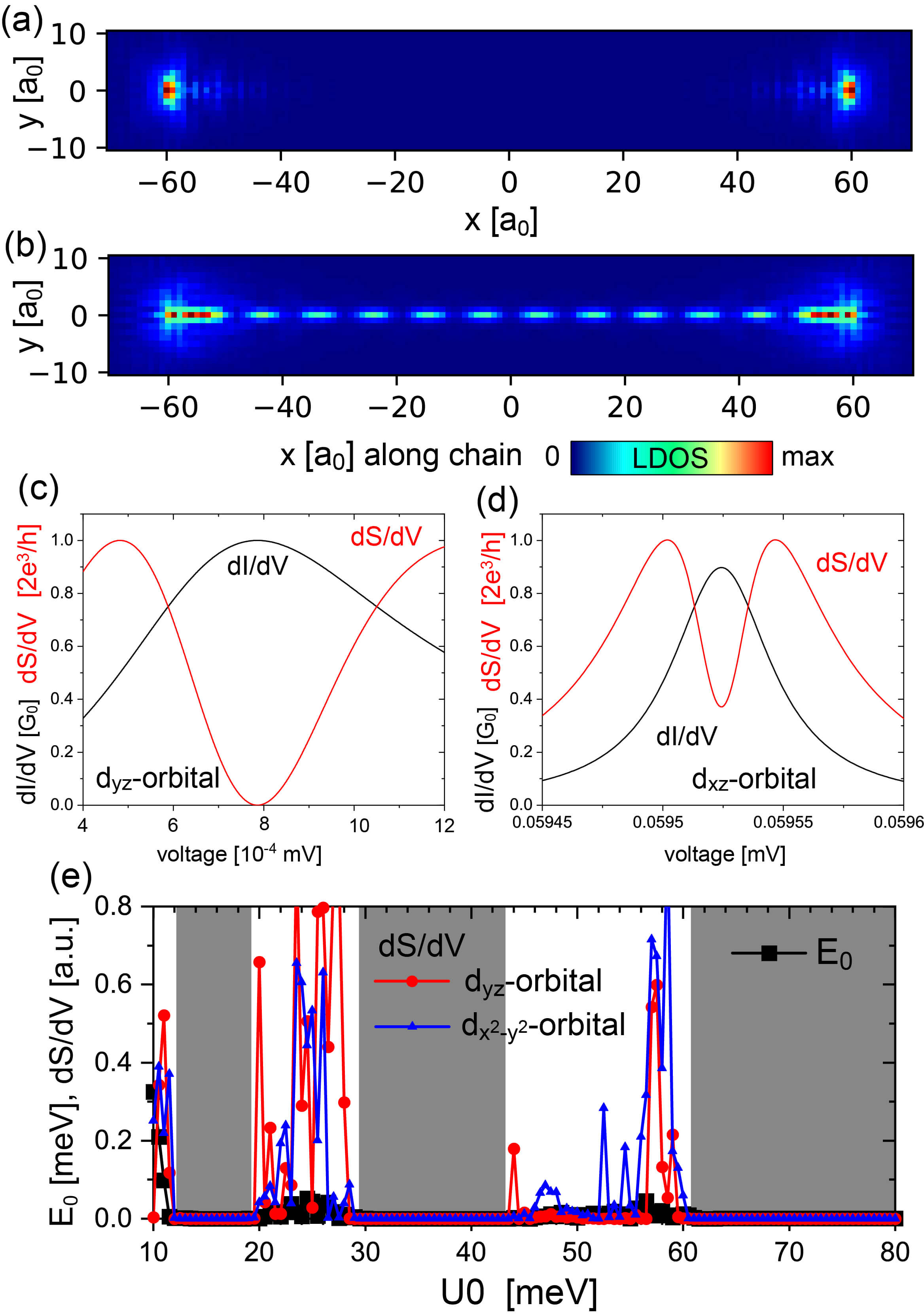}
\caption{Spatial structure of the LDOS around a line defect with $(\alpha, JS) = (7,7.5)$ meV for (a) the lowest energy state at $E_0=0.000786$, and (b) the next higher energy state at $E_1=0.0595$. The line defect has a length is $L=120a_0$ with scattering strength $U=72.5$ meV. (c), (d) The respective $dI/dV$ (black line) and $dS/dV$ (red line) for tunneling into the $d_{xy}$-orbital near energies corresponding to panels (a) and (b), respectively. (e) The energy of the lowest energy state, as well as $dS/dV$ for tunneling into the $d_{xy}$- and $d_{x^2-y^2}$-orbitals, as a function of $U_0$. }
\label{fig:Fig3}
\end{figure}
As previously shown \cite{Mascot2021}, when the underlying system is in a topological phase, there exist certain ranges of the scattering potential $U_0$ in which localized zero energy states emerge at the end of the line defect, whose spatial LDOS structure is consistent with that of MZMs, as shown in Fig.~\ref{fig:Fig3}(a). In contrast, the next higher energy state is delocalized along the chain, as shown in Fig.~\ref{fig:Fig3}(b). This qualitative difference between these two states is also reflected in the form of $dI/dV$ and $dS/dV$. For tunneling into the lowest energy state, we again find that when $dI/dV$ reaches $G_0$, $dS/dV$ vanishes, as shown in Fig.~\ref{fig:Fig3}(c), establishing that this state is a MZM. This result holds for tunneling into all of the 5 Fe $d$-orbitals (see SM Sec.~4). In contrast, for the next higher energy state [see Fig.~\ref{fig:Fig3}(d)], $dI/dV$ does not reach $G_0$, and $dS/dV$ does not vanish, implying that this state is not topological in nature. Since $dS/dV$ vanishes at the energy of a Majorana mode, a measurement of $dS/dV$ can be employed to identify topological phase transitions. For example, consider the evolution of the lowest energy state, $E_0$, associated with the line defect as a function of the scattering strength $U_0$ shown in Fig.~\ref{fig:Fig3}(e). For those regions of $U_0$ with $E_0 = 0$ (shown with a gray background), we find that the differential shot noise for tunneling into the $d_{yz}$- and  $d_{x^2-y^2}$-orbitals also vanishes. In contrast, already for small deviations of $E_0$ from zero, on the order of 1\% of the superconducting gap size, $dS/dV$ exhibits significant deviations from zero, indicating that the lowest energy state is no longer topological. Thus, $dS/dV$ represent a sensitive probe for the topological nature of very low energy states.

{\it Conclusions}
In conclusion, we have studied the form of the differential noise, $dS/dV$, and the differential conductance, $dI/dV$, for tunneling into localized MZMs in vortex cores and at the end of line defects,  and into chiral Majorana edge modes in \FSTT. Using a 5-band model that was previously extracted from fits to ARPES and STS experiments \cite{Sarkar2017}, we demonstrated that for all three cases, when the differential conductance, $dI/dV$ reaches the quantum of conductance, the differential shot-noise, $dS/dV$, as measured via scanning tunneling shot noise spectroscopy, vanishes. In contrast, for low energy states that are trivial, $dS/dV$ remains finite. These results demonstrate that $dS/dV$ is a sensitive probe for the topological nature of low-energy states, being able to discriminate between topological and trivial low-energy states. This ability, in turn, can be employed to detect topological phase transitions. These results hold for tunneling into any of the Fe $d$-orbitals of \FSTT, demonstrating that they are independent of the material's specific complex electronic bandstructure. In addition, we showed that a non-zero net supercurrent flows along domain walls separating topological and trivial regions. Thus, the combination of differential shot-noise, differential conductance, and measurements of local supercurrents via scanning probe microscopy techniques can provide a unique fingerprint for the existence of topological Majorana modes.

\noindent{\bf Acknowledgements}\\
We would like to thank S. Rachel for stimulating discussions. The theoretical work on supercurrents and the differential conductance near domain walls was supported by the U. S. Department of Energy, Office of Science, Basic Energy Sciences, under Award No. DE-FG02-05ER46225 (E.M and D.K.M.). The theoretical work on the differential shot noise was supported by the Center for Quantum Sensing and Quantum Materials, an Energy Frontier Research Center funded by the U. S. Department of Energy, Office of Science, Basic Energy Sciences under Award DE-SC0021238 (K.H.W and D.K.M.). V.M and D.J.V.H acknowledge support by the Center for Quantum Sensing and Quantum Materials, an Energy Frontier Research Center funded by the U. S. Department of Energy, Office of Science, Basic Energy Sciences under Award DE-SC0021238.

\end{document}

\title{Supplemental Material for \\Shot-noise and differential conductance as signatures of putative topological superconductivity in FeSe$_{0.45}$Te$_{0.55}$}

\author{Ka Ho Wong$^{1}$, Eric Mascot$^{1}$, Vidya Madhavan$^{2}$, Dale J. Van Harlingen$^{2}$, and Dirk K. Morr$^{1}$}
\affiliation{$^{1}$ University of Illinois at Chicago, Chicago, IL 60607, USA \\
$^{2}$ University of Illinois at Urbana Champaign, Champaign, IL 61801, USA}

\maketitle

\section{Theoretical Formalism}

To investigate the differential conductance, $dI/dV$, and the differential noise, $dS/dV$ for tunneling into Majorana modes, we consider the Hamiltonian that was proposed in Ref.\cite{Mascot2021} to explain the emergence of topological surface superconductivity in \FSTT. To compute $dI/dV$ and $dS/dV$, we need to a term, $H_t$ to the Hamiltonian that describes the electron tunneling from the STS tip into the system. The total Hamiltonian in real space is then given by $H = H_0 + H_{tip}$, where
\begin{align}
 H_{0} =& -\sum_{a,b=1}^{5} \sum_{{\bf r},{\bf r'},\sigma} t^{ab}_{\bf r,r'} c_{{\bf r},a,\sigma}^{\dagger} c_{{\bf r'},b,\sigma} - \sum_{a=1}^{5} \sum_{{\bf r},,\sigma} \mu_{aa}  c_{{\bf r},a,\sigma}^{\dagger} c_{{\bf r},a,\sigma} \nonumber \\
 & +i \alpha \sum_{a=1}^{5} \sum_{{\bf r}, {\bm \delta}, \sigma, \sigma'} c_{{\bf r},a,\sigma}^\dag \left( {\bm \delta}
\times {\bm \sigma}\right)^z_{\sigma \sigma'} c^\pd_{{\bf r}+{\bm \delta},a, \sigma'}
 + J  \sum_{a=1}^{5} \sum_{{\bf r},\sigma, \sigma'} {\bf S}_{\bf r} \cdot c_{{\bf r},a, \sigma}^\dag {\boldsymbol \sigma}_{\sigma \sigma'} c^\pd_{{\bf r},a, \sigma'}  \nonumber \\
& + \sum_{a=1}^{5}  \sum_{\langle \langle {\bf r},{\bf r'} \rangle \rangle} \Delta^{aa}_{{\bf r}{\bf r}'} c_{{\bf r},a,\uparrow}^{\dagger} c_{{\bf r'},a,\downarrow}^{\dagger} + {\rm H.c.}  \nonumber \\
H_{t} =& - t_0 \sum_{\sigma} \left( c^\dagger_{{\bf r}, a, \sigma} d _{\sigma} + H.c. \right)
 \label{eq:Hamiltonian}
\end{align}
Here $a,b=1,...,5$ are the orbital indices corresponding to the $d_{xz}$-, $d_{yz}$-, $d_{x^2-y^2}$-, $d_{xy}$-, and $d_{3z^2-r^2}$-orbitals, respectively,  $-t^{ab}_{\bf rr'} $ represents the electronic hopping amplitude between orbital $a$ at site ${\bf r}$ and orbital $b$ at site ${\bf r'}$ on a two-dimensional square lattice, $\mu_{aa}$ is the on-site energy in orbital $a$, $c_{{\bf r},a,\sigma}^{\dagger} (c_{{\bf r},a,\sigma})$ creates (annihilates) an electron with spin $\sigma$ at site ${\bf r}$ in orbital $a$, and ${\bm \sigma}$ is the vector of spin Pauli matrices. The superconducting order parameter $\Delta^{aa}_{{\bf r}{\bf r}'}$ represents intra-orbital pairing between next-nearest neighbor Fe sites ${\bf r}$ and ${\bf r}'$ (in the 1 Fe unit cell), yielding a superconducting $s_\pm$-wave symmetry \cite{Sarkar2017}. Moreover, $\alpha$ denotes the Rashba spin-orbit interaction arising from the breaking of the inversion symmetry at the surface\,\cite{NadjPerge2014} with $\bm \delta$ being the vector connecting nearest neighbor sites. Moreover, $-t_0$ represents the tunneling elements for electron tunneling from the STS tip into orbital $a$ at site ${\bf r}$.

To compute $dI/dV$ and $dS/dV$, we need to find the Greens functions of the entire system, superconductor plus STS tip. To this end, we introduce the following spinor for the superconductor
\begin{align}
\Psi^\dagger &= (\psi^\dagger_1, \psi^\dagger_2, ...., \psi^\dagger_N ) \nonumber \\
\psi^\dagger_p & = \left( \psi^\dagger_{p,\uparrow}, \psi^\dagger_{p,\downarrow}, \psi^T_{p,\downarrow}, -\psi^T_{p,\uparrow} \right) \nonumber \\
\psi^\dagger_{p,\sigma} & = \left( c_{p,\sigma,xz }^{\dagger}, c_{p,\sigma,yz}^{\dagger}, c_{p,\sigma,x^2-y^2 }^{\dagger}, c_{p,\sigma,xy }^{\dagger}, c_{p,\sigma,3z^2-r^2 }^{\dagger}\right)
\end{align}
where $1,2, ...$ denotes a site in the system, with $N=N_x \times N_y$ being the total number of sites. The Hamiltonian $H_0$ in Eq.(\ref{eq:Hamiltonian}) can then be written in the form
\begin{align}
  H_0 &= \Psi^\dagger {\hat H}_0 \Psi
\end{align}
and we can define the Greens function matrix of the superconductor (decoupled from the STS tip) in Matsubara time via
\begin{align}\label{eq:GF}
  {\hat g}_{sc}(\tau) & = -\langle {\mathcal T}_\tau \Psi(\tau) \Psi^\dagger(0) \rangle \ .
\end{align}
yielding for ${\hat g}_{sc}$ in Matsubara frequency space
\begin{align}
  {\hat g}_{sc}(i\omega_n) & = \left[ i\omega_n {\hat 1} - {\hat H}_0 \right]^{-1}  \ .
  \label{eq:MatsGF}
\end{align}
Similarly, we can define the Greens function of the tip in Matsubara time via
\begin{align}\label{eq:GFtip}
  {\hat g}_{\rm tip}(\tau) & = -\langle {\mathcal T}_\tau \phi(\tau) \phi^\dagger(0) \rangle
\end{align}
where
\begin{align}
  \phi^\dagger & = \left( d^\dagger_{\uparrow}, d^\dagger_{\downarrow}, d_{\downarrow}, -d_{\uparrow} \right)
\end{align}

Next, we employ the Dyson equation to include the tunneling between the STS tip and the superconductor, yielding the full Greens function of the entire system via
\begin{align}
	G^{<,>}(\omega)&=[\hat{1}-{\hat g}^{r}(\omega){\hat H}_t]^{-1}{\hat g}^{<,>}(\omega)[{\hat 1}-{\hat H}_t {\hat g}^{a}(\omega)]^{-1}\nonumber \\
	G^{r,a}(\omega)&=[({\hat g}^{r,a}\omega))^{-1}-{\hat H}_{t}]^{-1}
\label{eq:fullGF}
\end{align}
where, with $x=<,>,r,a$
\begin{align}
  {\hat g}^{x}(\omega) & = \left(
\begin{array}{cc}
\hat{g}_{{\rm tip}}^{x}(\omega) & 0  \\
0 & \hat{g}_{\rm sc}^{x}(\omega)
\end{array}%
\right)
\end{align}
$\hat{g}_{\rm sc}^{r,a}(\omega)$ are the retarded and advanced forms of Eq.(\ref{eq:MatsGF}), and $\hat{g}_{\rm sc}^{<}(\omega)$ is the lesser Greens function matrix given by
\begin{align}
{\hat g}_{sc}^<(\omega) &= -n_F(\omega)({\hat g}_{sc}^r(\omega)-{\hat g}_{sc}^a(\omega))
\end{align}
where $n_F(\omega)$ is the Fermi distribution function. For the tip, we consider the wide-band limit, such that
\begin{align}
\hat{g}_{{\rm tip}}^{r,a} & = \mp i \pi N_0 {\hat 1} \nonumber \\
\hat{g}_{{\rm tip}}^{<} &  = i \pi N_0 \left(
\begin{array}{cccc}
n_F(\omega-eV) & 0 & 0 & 0  \\
0 & n_F(\omega-eV) & 0 & 0 \\
0 & 0 & n_F(\omega+eV) & 0 \\
0 & 0 & 0 & n_F(\omega+eV)
\end{array}%
\right)
\end{align}
where $e$ is the electron charge, and $V$ is the potential difference between the tip and the grounded superconductor. Moreover, using the expanded spinor
\begin{align}
  {\bar \Psi}^\dagger &= (\phi, \psi^\dagger_1, \psi^\dagger_2, ...., \psi^\dagger_N )
\end{align}
we can rewrite $H_{t}$ as
\begin{align}
  H_{t} = {\bar \Psi}^\dagger {\hat H}_t {\bar \Psi}
\end{align}
which defines ${\hat H}_{t}$ in Eq.(\ref{eq:fullGF}).

The current flowing between the STS tip and the superconductor is given by  \cite{Rachel2017}
\begin{align}
I(V)=\frac{2 e t_0}{h}\sum_{\sigma=\uparrow,\downarrow}
\int_{-eV}^{eV}\dd{\varepsilon}\mathrm{Re}
\left[G^{<}_{dc}\left(\vb{r},\sigma,\sigma,\varepsilon\right)\right]
\label{eq:IV}
\end{align}
where $\omega = eV/\hbar$.  The zero-frequency shot-noise for tunneling into orbital $a$ at site ${\bf r}$ is defined as the current-current correlation function in real time
\begin{equation}
S({\bf r}, a, t,t^\prime)=\langle \left\{ {\hat I_a}({\bf r},t)-I_a({\bf r}),{\hat I}({\bf r},t^\prime)-I_a({\bf r}) \right\} \rangle
\end{equation}
where ${\hat I_a}$ is the current operator, and $I_a({\bf r})$ is the steady-state current that flows from the tip into orbital $a$ at site ${\bf r}$. We then obtain
\begin{align}
S_0({\bf r}, a, \omega = 0) &=\frac{2e^2 t_0^2}{h}	\sum_{\sigma,\sigma'=\uparrow,\downarrow}\int_{-eV}^{eV}\dd{\varepsilon}
\left[G_{tt}^{>}\qty(\sigma;\sigma',\varepsilon)G^{<}_{ss}\qty({\bf r}, a,\sigma',\sigma,\varepsilon)
+G_{cc}^{>}\qty({\bf r}, a,\sigma,\sigma',\varepsilon)G^{<}_{dd}\qty(\sigma',\sigma,\varepsilon) \right. \nonumber \\
&  + \left[ F^{>}_{cc}\qty({\bf r}, a, \sigma,\sigma',\varepsilon)\right]^* F^{<}_{dd}\qty(\sigma',\sigma,\varepsilon)+\left[F^{>}_{dd}\qty(\sigma,\sigma',\varepsilon)\right]^* F^{<}_{cc}\qty({\bf r}, a,\sigma',\sigma,\varepsilon) \nonumber\\
& -G^{>}_{dc}\qty({\bf r},a, \sigma,\sigma',\varepsilon)G^{<}_{dc}\qty({\bf r},a,\sigma',\sigma,\varepsilon)-G^{>}_{cd}\qty({\bf r},a,\sigma,\sigma',\varepsilon)G^{<}_{cd}\qty({\bf r},a,\sigma',\sigma,\varepsilon) \nonumber\\
&  \left. -\left[F^{>}_{cd}\qty({\bf r},a,\sigma,\sigma',\varepsilon)\right]^*F^{<}_{cd}\qty({\bf r},a,\sigma',\sigma,\varepsilon)-\left[F^{>}_{dc}\qty({\bf r},a,\sigma,\sigma',\varepsilon)\right]^*F^{<}_{dc}\qty({\bf r},a,\sigma',\sigma,\varepsilon) \right]	\ .
\label{eq:SV}
\end{align}
All Greens function in Eqs.(\ref{eq:IV}) and (\ref{eq:SV}) are elements of the Greens function matrices given in Eq.(\ref{eq:fullGF}), with $G$ and $F$ corresponding to the normal and anomalous Greens functions, $t,s$ denoting the tip and the system, and all Green's functions involving the system are evaluated at site ${\bf r}$ and for orbital $a$. For example, $G^{<}_{dc}({\bf r},a,\sigma',\sigma,\varepsilon)$ is the lesser form of the Greens function in Matsubara time given by
\begin{align}
G_{dc}(\sigma'; {\bf r},a,\sigma,\tau) = -\langle {\cal T}_\tau d_{\sigma^\prime}(\tau) c^\dagger_{{\bf r},a,\sigma}(0) \rangle
\end{align}

\section{$dS/dV$ and $dI/dV$ for tunneling into a chiral Majorana edge mode}

\begin{figure}[t]
\centering
\includegraphics[width=16cm]{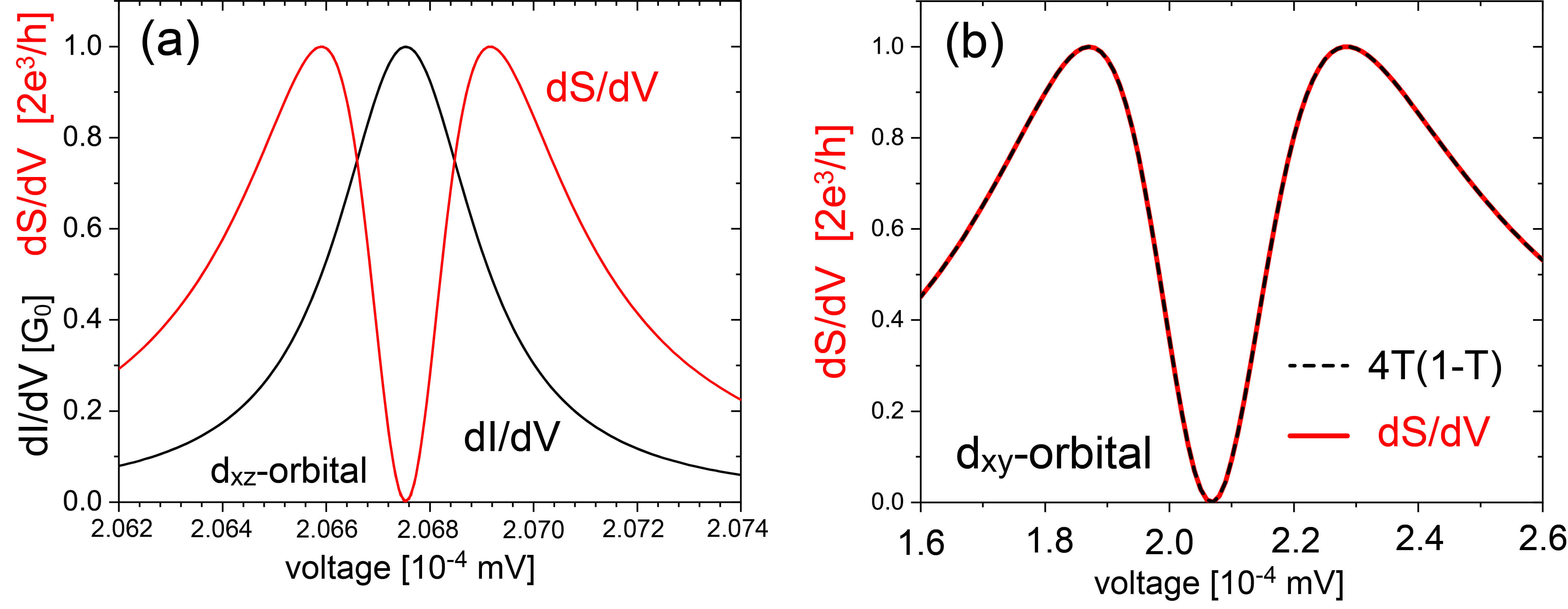}
\caption{Domain wall for $\alpha=7$ meV separating a topological $C=1$ regions with $J=7.5$ meV from a trivial $C=0$ region with $J=0$. (a) $dS/dV$ and $dI/dV$ for the lowest energy Majorana state for tunneling into the $d_{xz}$-orbital. (b) Comparison of $dS/dV$ (red line) and $T(V)[1-T(V)]$ (dashed black line) for the lowest energy Majorana state and tunneling into the $d_{xy}$-orbital.}
\label{fig:SI_Fig1}
\end{figure}
For the domain wall discussed in Fig.~1 of the main text, only the $d_{xy}$- and $d_{xz}$-orbitals possess a non-vanishing spectral weight near zero energy at the location of the domain wall. The resulting $dI/dV$ and $dS/dV$ for tunneling into the $d_{xy}$-orbital are shown in Fig.~1(d) of the main text, while those for tunneling into the $d_{xz}$-orbital are shown in Fig.~\ref{fig:SI_Fig1}(a). Similar to the results shown in Fig.~1(d), we find that $dS/dV$ vanishes when $dI/dV$ reaches $G_0$. Since the spectral weight of the $d_{xz}$-orbital is significantly smaller than that of the $d_{xy}$-orbital at the domain wall, we find that the energy width of $dI/dV$ for the $d_{xz}$-orbital is significantly smaller than that of the $d_{xy}$-orbital. The spectral weight of the chiral Majorana modes in the remaining three orbitals is vanishingly small at the domain wall, such that they are not considered here.

To demonstrate the relation between $dI/dV$ and $dS/dV$, we write
\begin{align}\label{eq:dIdV_T}
  \frac{dI(V)}{dV} = G_0 T(V)
\end{align}
where $T(V)$ is the bias-dependent transmission coefficient. We then find that $dS/dV$ can be written as
\begin{align}\label{eq:dSdV_T}
  \frac{dS(V)}{dV} = \frac{8e^3}{h} T(V) [1-T(V)]
\end{align}
In Fig.~\ref{fig:SI_Fig2}(b), we plot $dS/dV$ (in units of $2 e^3/h$) together with $T(V) [1-T(V)]$, as extracted from Eq.(\ref{eq:dIdV_T}). The very good agreement between these two results demonstrates the functional relationship between $dI/dV$ and $dS/dV$ shown in Eqs.(\ref{eq:dIdV_T}) and (\ref{eq:dSdV_T}).

\section{$dS/dV$ and $dI/dV$ for tunneling into a localized MZM in a vortex core}

\begin{figure*}[t]
\centering
\includegraphics[width=16cm]{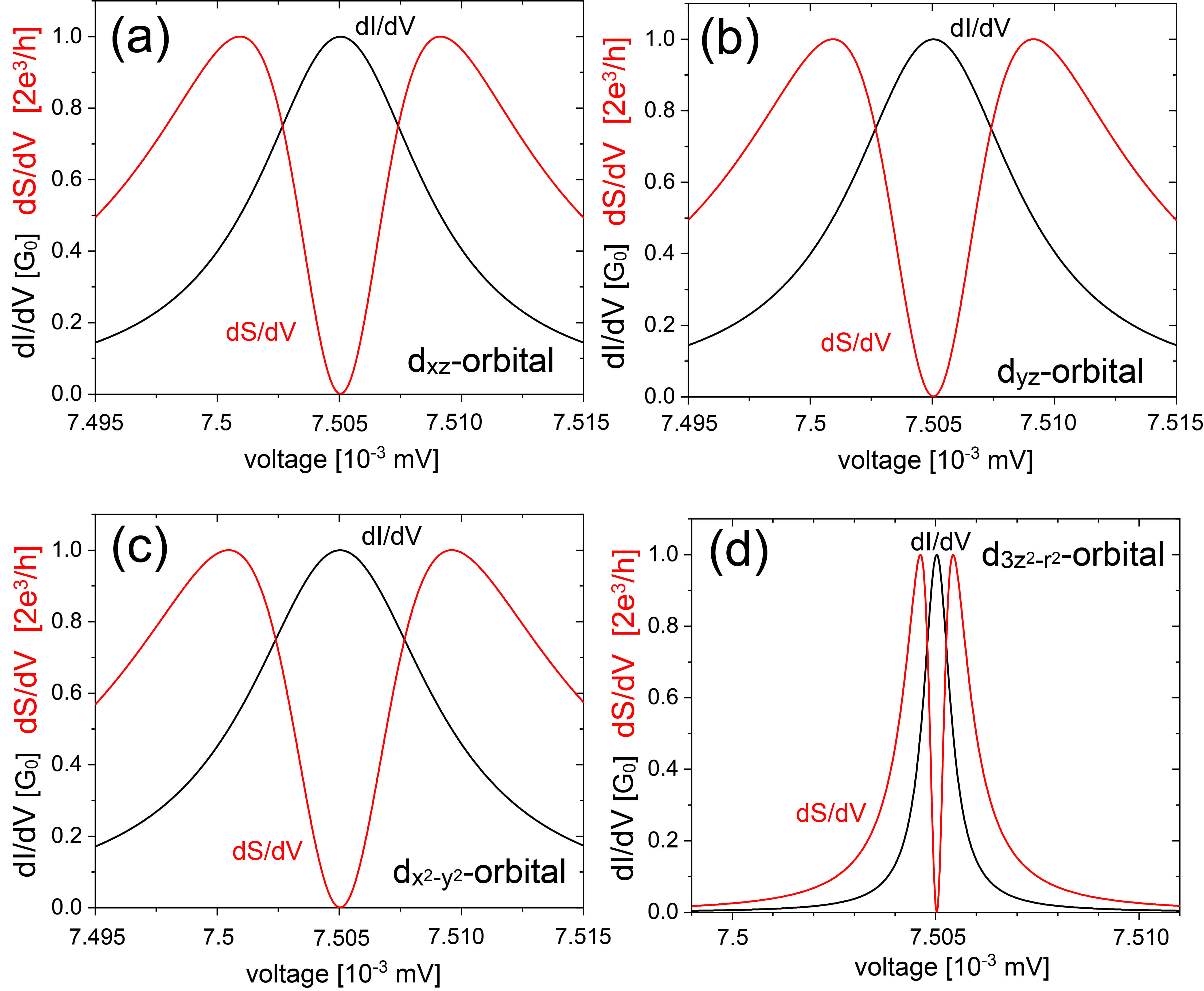}
\caption{$dS/dV$ and $dI/dV$ for the lowest energy Majorana state for tunneling into the (a) $d_{xz}$-orbital, (b) $d_{yz}$-orbital, (c) $d_{x^2-y^2}$-orbital, and (d) $d_{3z^2-r^2}$-orbital of a MZM located at the center of a vortex core in the topological $C=1$ phase with $(\alpha,J)=(7,8)$ meV for a uniform magnetic flux of $1$ Wb.}
\label{fig:SI_Fig2}
\end{figure*}

$dS/dV$ and $dI/dV$ for tunneling into the $d_{xy}$-orbital of a localized MZM in a vortex core was shown in Fig.~2 of the main text. In Fig.~\ref{fig:SI_Fig2}, we present $dS/dV$ and $dI/dV$ for tunneling into the four remaining orbitals. In each case, we find that $dS/dV$ vanishes when $dI/dV$ reaches the quantum of conductance, $G_0$. This result demonstrates that the vanishing of $dS/dV$ is independent of the complex electronic structure of \FSTT.

\section{$dS/dV$ and $dI/dV$ for tunneling into a localized MZM at the end of a line defect}

\begin{figure*}[t]
\centering
\includegraphics[width=16cm]{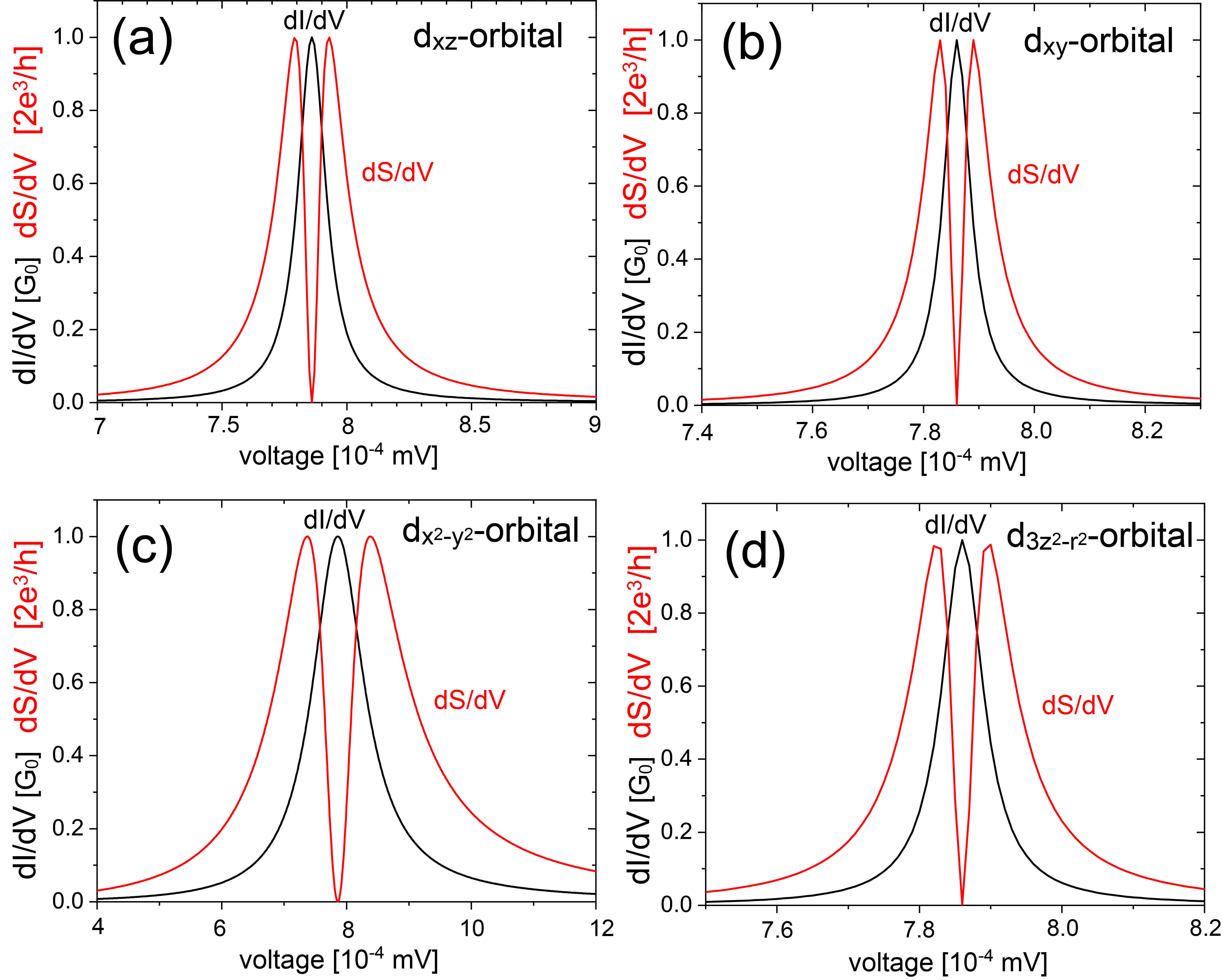}
\caption{$dS/dV$ and $dI/dV$ for the lowest energy Majorana state for tunneling into the (a) $d_{xz}$-orbital, (b) $d_{xy}$-orbital, (c) $d_{x^2-y^2}$-orbital, and (d) $d_{3z^2-r^2}$-orbital of a MZM located at the end of a line defect of length $L=120a_0$ with $(\alpha, JS) = (7,7.5)$ meV.}
\label{fig:SI_Fig3}
\end{figure*}

$dS/dV$ and $dI/dV$ for tunneling into the $d_{xz}$-orbital of a localized MZM at the end of a line defect was shown in Fig.~3(c) of the main text. Similar to those results, we find that for tunneling into the four remaining orbitals (see Fig.~\ref{fig:SI_Fig3}), $dS/dV$ vanishes when and $dI/dV$ reaches the quantum of conductance.

\bibliography{Topo_FeSeTe_noise}